\documentclass[aps,twocolumn]{revtex4}
\usepackage[dvips]{graphics,graphicx}
\usepackage{amsmath}
\begin{document}

\title{On quantum phase transitions in tilted 2D lattices}
\author{Andrey~R.~Kolovsky$^{1,2}$}
\affiliation{$^1$Kirensky Institute of Physics, 660036 Krasnoyarsk, Russia}
\affiliation{$^2$Siberian Federal University, 660041 Krasnoyarsk, Russia}
\date{\today}
\begin{abstract}
We discuss the quantum phase transition from the Mott-insulator state to the density-wave state for cold Bose atoms in a 2D square lattice as the lattice is adiabatically tilted along one of its primary axes. It is shown that a small misalignment of the tilt drastically changes the result of the adiabatic passage and, instead of the density-wave state, one obtains a disordered state. Intrinsic relation of the problem to Bloch oscillations of hard-core bosons in a 2D lattice is illuminated.  
\end{abstract}
\maketitle

{\em Introduction.}
Tilted 1D optical lattices are the standard setup for studying Bloch oscillations (BOs) of non-interacting and interacting cold atoms \cite{Daha96,Mors01,57,61,80,Poli11,Mein14a}, with important applications to precision measurements of the gravitational force \cite{Poli11} and  interatomic interaction constant \cite{Mein14a}. A different direction of research is quantum phase transitions in tilted lattices \cite{Sach02,Simo11}. (We also mention relevant studies of the quench dynamics \cite{Seng04,64,Rubb11,Kolo12,Mein13,Mein14b}.) These are rather specific phase transitions because, formally, the system has no ground state. Nevertheless, by adiabatically tilting the lattice one observes continuos evolution of the Mott-insulator (MI) state of Bose atoms into the density-wave (DW) state \cite{Simo11}. As it was explained in Ref.~\cite{Sach02} by mapping the system of Bose atoms into an effective system of interacting spins, this `not-ground-state' transition corresponds to the common ground-state phase transition from ferromagnetic to anti-ferromagnetic ordering of the Heisenberg spin chain. 

More phases are expected if we consider Bose atoms in 2D lattices \cite{Piel11}.  In fact, two-dimensional systems offer a freedom in choosing the lattice geometry and orientation of a static force ${\bf F}$ relative to primary axes of the lattice. However, since we face not-ground-state transition, this freedom may lead to additional effects that are absent in the effective ground-state problem. In this work we discuss one of them, namely, an effect caused by the lattice misalignment. In more detail,  we shall analyze formation of the density-wave phase in a square lattice which is tilted in the $y$ direction with some uncertainty $F_x\ll F_y$. It will be shown that this small misalignment completely changes the result of the adiabatic passage and, instead of the ordered DW state, one obtains a disordered state.

{\em Phase transition in a ladder.}
To illustrate the role of the weak component $F_x$ we first consider the square lattice which consists of two rows, i.e., a two-leg ladder:
\begin{eqnarray}
\nonumber
\widehat{H}(t)= -\frac{J_x}{2}\sum_{m=1}^2 \sum_{l=1}^L  \left(\hat{a}^\dagger_{l+1,m}\hat{a}_{l,m}e^{-iF_x t} + h.c.\right) \\
\nonumber
-\frac{J_y}{2}\sum_{l=1}^L  \left(\hat{a}^\dagger_{l,2}\hat{a}_{l,1} + h.c.\right) +\frac{U}{2}\sum_{l,m} \hat{n}_{l,m}(\hat{n}_{l,m}-1) \\
\label{1}
- F_y\sum_l(\hat{n}_{l,2}-\hat{n}_{l,1})  \;.
\end{eqnarray}
In the Hamiltonian (\ref{1}) the last term is the lattice tilt in the $y$ direction and non-zero component $F_x$ is taken into account by using the gauge transformation.  We mention that the system (\ref{1}) is of its own interest because it can be realized experimentally by using double-periodic optical potential, in spirit of the recent experiment \cite{Atal14}. 
\begin{figure}
\center
\includegraphics[width=8cm,clip]{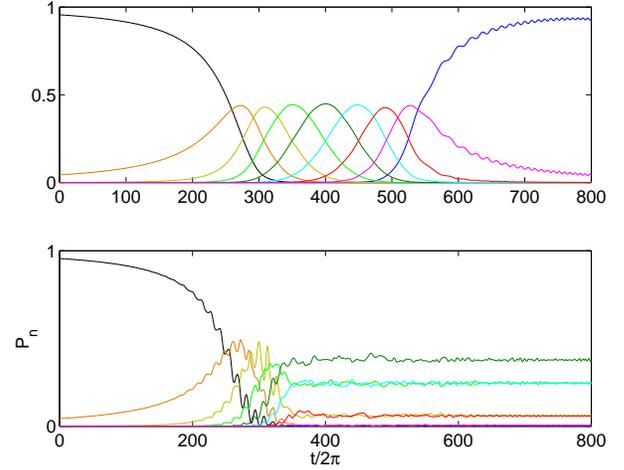}
\caption{Probabilities $P_n(t)$ to find $n$ doublons in the system for $F_x=0$, upper panel, and $F_x=0.01$, lower panel. The field component $F_y$ is increased linearly in time, where the depicted time interval corresponds to $0.8\le F_y \le 1.2$. The other parameters are $L=8$, $U=1$ and $J_x=J_y=0.02$.}
\label{fig1}
\end{figure}

Since we are interested in the case $U\gg J_x,J_y$, we can truncate the Hilbert space to the resonant subspace spanned by the Fock states
\begin{equation}
\label{2}
|{\bf n}\rangle=\left[
\begin{array}{cccc}
n_{1,2}&n_{2,2}&\ldots&n_{L,2}\\n_{1,1}&n_{2,1}&\ldots&n_{L,1} 
\end{array}
\right] \;,
\end{equation}
where occupation numbers $n_{l,m}$ may take value 1 or 0 if $m=1$, and 1 or 2 if $m=2$. Then the discussed phase transition corresponds to evolution of the MI state, where $n_{l,m}=1$, into the ordered doublon state,  where $n_{l,1}=0$ and $n_{l,2}=2$. This evolution is illustrated in Fig.~\ref{fig1}(a) where $F_x=0$ and we increase $F_y$ linearly in time with the rate $\nu=0.0005$. Different curves are probabilities $P_n(t)$ to find $n$ doublons in the ladder at a given time.  It is seen in Fig.~\ref{fig1}(a) that  we obtain the DW state at the end of  the adiabatic passage.

This result changes drastically if $F_x\ne0$, see Fig.~\ref{fig1}(b). Now the final state of the system is a random state from the micro-canonical distribution, where probabilities $P_n$ are given by relative dimensions of the corresponding subspaces of the Hilbert space, 
\begin{equation}
\label{22}
P_n={\cal N}_n/{\cal N} \;.
\end{equation}
(For $L=8$ the total dimension of the Hilbert space ${\cal N}= 1620$ and ${\cal N}_n= 1, 8,100,392,618,392,100,8,1$, respectively.)  As it will be explained below, the physics behind this phenomenon is self-thermalization of the system due to BOs of the quasiparticles (doublons and holes) which are dynamically created when we tilt the ladder in the $y$ direction. Since quasiparticles behave as hard-core (HC) bosons, one gets useful insight in the problem by studying BOs of HC bosons in a ladder. 

{\em Bloch oscillations of HC bosons.}
Bloch dynamics of HC bosons in a ladder is governed by the time-dependent Hamiltonian
\begin{eqnarray}
\nonumber
\widehat{H}(t)= -\frac{J_x}{2}\sum_{m=1}^2 \sum_{l=1}^L  \left(\hat{b}^\dagger_{l+1,m}\hat{b}_{l,m}e^{-iFt} + h.c.\right) \\
\label{3}
-\frac{J_y}{2}\sum_{l=1}^L  \left(\hat{b}^\dagger_{l,2}\hat{b}_{l,1} + h.c.\right) \;,
\end{eqnarray}
where  $\hat{b}^\dagger_{l,m}$ and $\hat{b}_{l,m}$ are the hard-core creation and annihilation bosonic operators, $(\hat{b}^\dagger_{l,m})^2=0$, and $F\equiv F_x$ is a static field parallel to the ladder legs. We shall restrict ourselves by the filling factor $1/2$ because in this case the Hilbert space of the system (\ref{3}) and the system (\ref{1}) are isometric. In fact, let 
\begin{equation}
\label{4}
|{\bf n}\rangle=\left[
\begin{array}{cccc}
n_{1,2}&n_{2,2}&\ldots&n_{L,2}\\n_{1,1}&n_{2,1}&\ldots&n_{L,1} 
\end{array}
\right] \;,\quad n_{l,m}=0,1 \;,
\end{equation} 
is the complete set of Fock states of HC bosons at half-filling. Then the resonant subset of Fock states of the system (\ref{1}) is obtained by adding unity to every element in the upper row, leaving the low row unchanged. The isometric Hilbert spaces imply the same skeleton of the Hamiltonian matrix, although the values of non-zero matrix elements may differ by factor $\sqrt{2}$ due to bosonic enhancement of tunneling in the original problem.  

\begin{figure}
\center
\includegraphics[width=8cm,clip]{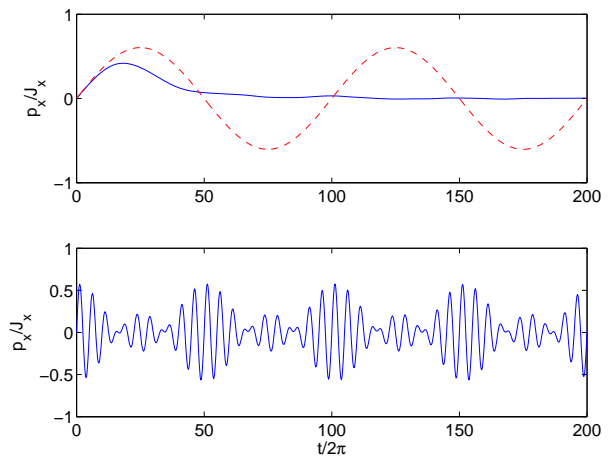}
\caption{Bloch oscillations of HC bosons in a ladder at half-filling. The mean momentum per particle is shown as the function of time. Parameters are $J_x=J_y=0.02$,  $L=N=8$, $F=0.01$ (upper panel) and $F=0.2$ (lower panel). Initial wave function is given by the ground state of the system for $F=0$. The dashed line in the upper panel is BOs of non-interacting fermions at half-filling.}
\label{fig4}
\end{figure}

It is well know that HC bosons in a 1D lattice can be mapped into the system of non-interacting fermions. This is, however, not the case for HC bosons in a ladder. Here any attempt of mapping leads to effective interactions and, thus, we face with BOs of interacting  particles. Previous studies of Bloch dynamics of interacting atoms revealed two qualitatively different regimes of BOs \cite{57,61,80,Mein14a}. These are the quasi-periodic  BOs, which take place for a strong field $F$, and decaying BOs, which is the case for a weak field. In the latter case the system gets thermalized, i.e., every Fock state becomes equally occupied. We found remarkable similarities between BOs of HC bosons in a ladder and BOs of weakly ($U\sim J$) interacting bosons in a 1D lattice. In particular, depending on the field strength, BOs of HC bosons in the ladder either irreversibly decay  or show a quasi-periodic dynamics, see Fig.~\ref{fig4}.  In the former case the irreversible decay indicates self-thermalization of the system. Using the above mentioned similarity with the original problem this explains the result (\ref{22}). 

{\em Infinite 2D lattices.}
The presented results suggest the following picture of phase transition in the 2D lattice. When we tilt the lattice in the $y$ direction to $F_y\approx U$ we produce particle-hole excitations of the MI state. These quasiparticles can move in the $x$ direction and, if $F_x\ne 0$, this motion causes self-thermalization of the system within the characteristic time $T_B=2\pi/F_x$.  To avoid this self-thermalization, the evolution time must be smaller than the Bloch period $T_B$.  On  the other hand, to insure adiabatic passage, this time should be as large as possible. These contradicting conditions introduce severe restriction on the lattice misalignment, which we shall discussed in some more details.

\begin{figure}
\center
\includegraphics[width=8cm,clip]{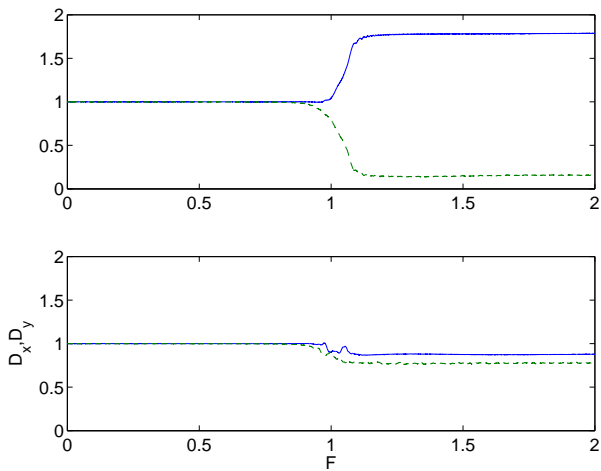}
\caption{Order parameters $D_x(t)$, blue solid lines, and $D_y(t)$, green dashed lines, for $F_x/F_y=0$, upper panel,  and $F_x/F_y=0.01/\sqrt{1-0.01^2}$, lower panel. The static field is increased linearly in time with the rate $\nu=10^{-3}$, the lattice size is $3\times4$.}
\label{fig5}
\end{figure}

We mimic phase transition in an infinite lattice by considering finite lattices up to $4\times4$ sites and imposing periodic boundary conditions in both the $x$ and $y$ directions. To characterize the final state we introduce the order parameters
\begin{eqnarray}
\label{5}
D_x(t)=N^{-1}\langle\psi(t)|\sum_{l,m} \hat{n}_{l+1,m}\hat{n}_{l,m} |\psi(t)\rangle \;, \\
\label{6}
D_y(t)=N^{-1}\langle\psi(t)|\sum_{l,m} \hat{n}_{l,m+1}\hat{n}_{l,m} |\psi(t)\rangle \;,
\end{eqnarray}
where $N$ is number of atoms coinciding with number of sites.  It is easy to prove that the MI state corresponds to $D_x=D_y=1$ while the DW state has $D_x=2$ and $D_y=0$. For $F_x=0$ (i.e., the precise alignment) and the rate $\nu=0.001$ dynamics of the order parameters (\ref{5},\ref{6}) is depicted in Fig.~\ref{fig5}.  First of all we notice that the final state deviates from the ideal DW state even for $F_x=0$. This is a consequence of the high-order resonant tunneling which happens at $F=U/j$ where $j$ is an integer number (see the recent works \cite{Mein13,Mein14b} and references therein). For the considered rate $\nu=10^{-3}$ only the second-order process is important. It creates small number of doublons and holes in the next to the nearest rows of the lattice as $F$  is increased above $U/2$. For the subsequent first-order process at $F\approx U$ these objects play the role of difects which prohibit particle-hole excitations of the MI state in their vicinity. For this reason the total number of doublons never reaches the maximally possible number $N/2$ and the order parameters deviate from their extreme values.

Small deviation of the final state from the DW state due to the high-order resonant tunneling is a minor effect in comparison with the effect of the lattice alignment.  The lower panel in Fig.~\ref{fig5} shows the result of numerical simulations for $F_x/F_y\approx 0.01$. It is seen that the final state practically has no correlations. At the same time, the total number of doublons remains pretty high (results are not shown). Thus  we end up with a disordered state of doublons.

{\em Conclusion.}
We analyzed response of the Mott-insulator state of cold atoms in a square 2D optical lattice to a static field ${\bf F}$ which is adiabatically increased from zero to a value above the interaction energy $U$. If the field ${\bf F}$ is precisely aligned with the $y$ axis of the lattice the Mott-insulator state was shown to evolve in the density-wave state where every second row is empty and the rest rows are filled with doublons (two atoms in one site). This result can be viewed as two-dimensional generalization of the quantum phase transition observed in tilted 1D optical lattices \cite{Simo11}. The new effect was found if the static field is slightly misaligned with respect to the $y$ axis. In this case the final state of the system is a disordered state of doublons and holes with vanishing correlations.  The physics behind this effect proved to be self-thermalization of the system due to Bloch oscillations of the quasiparticles (doublons and holes) in the $x$ direction. 

In the present work we restricted ourselves by considering the field orientation close to the primary $y$ axis of the square 2D lattice. Obviously, all reported results hold true in the situation where ${\bf F}$ is close the $x$ axis.  The case of other orientations, for example $F_x/F_y\approx 1$, is more involved and is expected to strongly depend on the lattice geometry. In fact, the simple square lattice is a rare exclusion where single-particle wave functions, known as the Wannier-Stark states, are localized for any orientations of the static field except those coinciding with primary axes. In a generic  2D lattice the quantum particle is delocalized in the direction orthogonal ${\bf F}$ for every `rational' orientation of the static field which is given by arbitrary superposition of the translation vectors  \cite{101}. Thus one may expect a similar result: we shall observe transition to an ordered state if the vector ${\bf F}$ points from one lattice site to a nearby lattice site exactly and self-thermalization of the system if ${\bf F}$ slightly deviates from this direction. The detailed analysis of the outlined problem will be presented elsewhere.

{\em Acknoledgements.}
The author acknowledges fruitful discussions with M.~Fleischhauer and  F.~Grusdt, and hospitality of the University of Kaiserslautern, where a part of this work was conducted.


\end{document}